\documentstyle[preprint,aps,psfig]{revtex}
\newcommand{\beq}{\begin{equation}}
\newcommand{\eeq}{\end{equation}}
\newcommand{\beqa}{\begin{eqnarray}}
\newcommand{\eeqa}{\end{eqnarray}}
\newcommand{\ba}{\begin{array}}
\newcommand{\ea}{\end{array}}

\begin{document}
\draft


\widetext 
\title{Critical Temperature of an Interacting Bose Gas \\
in a Generic Power-Law Potential} 
\author{Luca Salasnich} 
\address{Istituto Nazionale per la Fisica della Materia, 
Unit\`a di Milano, \\
Dipartimento di Fisica, Universit\`a di Milano, \\
Via Celoria 16, 20133 Milano, Italy\\
E-mail: salasnich@mi.infm.it}
\maketitle
\begin{abstract}
We investigate the critical temperature 
of an interacting Bose gas confined in a trap described by a 
generic isotropic power-law potential. We compare the results 
with respect to the non-interacting case. In particular, 
we derive an analytical formula for the shift of the critical temperature 
holding to first order in the scattering length. 
We show that this shift scales as $N^{n\over 3(n+2)}$, 
where $N$ is the number of Bosons and $n$ is the exponent 
of the power-law potential. Moreover, the sign of the shift 
critically depends on the power-law exponent $n$. 
Finally, we find that the shift of the critical temperature 
due to finite-size effects vanishes as $N^{-{2n\over 3(n+2)}}$. 
\end{abstract}
\vskip 0.5cm 


\newpage 

\narrowtext

Nowadays more than twenty experimental groups have achieved 
Bose-Einstein condensation (BEC) in trapped 
alkali-metal atoms by using different geometries of the confining trap 
and atomic species.$^{1-3}$ The experimental results on these dilute 
gases of weakly-interacting Bosons are in quite good agreement 
with the theoretical predictions both above and below the 
BEC critical temperature (for a recent review see Ref. 4). 
\par 
An interesting problem is the behavior of the 
critical temperature $T_c$ as a function of the number $N$ 
of Bosons. Unfortunately, the experimental data 
of the critical temperature are not 
very precise but they seem to qualitatively confirm the 
predictions. In many experiments with alkali-metal 
atoms the trap can be described by a harmonic potential 
$U({\bf r})=(1/2)m\omega^2 r^2$. 
In this case, it is well known that in the large $N$ limit 
the critical temperature $T_c^{(0)}$ 
of an ideal gas scales as $N^{1\over 3}$,$^{5}$ 
while the shifts of the critical temperature 
due to the finite number of Bosons and the interaction 
scale as $N^{-{1\over 3}}$ and $N^{1\over 6}$, 
respectively.$^{6,7}$  
Note that the shape of the trapping 
potential plays a decisive role for the critical 
temperature, as we have recently shown with a dilute gas of 
hydrogen atoms in a Ioffe trap.$^{8}$ 
\par 
An important class of trapping potentials is 
provided by power-law potentials $U({\bf r})=A \; r^n$. 
Power-law potentials have been proposed to cool the 
Bose gas in a reversible 
way by adiabatically changing the shape of the trap 
at a rate slow compared to the internal equilibration 
rate.$^{9-11}$  For these potentials 
it has been shown that the critical temperature $T_c^{(0)}$ 
of an ideal Bose gas scales as $N^{2n\over 3(n+2)}$ 
in a 3-dimensional space$^{5}$ and as $N^{2n\over d(n+2)}$ 
in a generic d-dimensional space.$^{12}$  
\par 
In this paper we extend previous results of 
an ideal Bose gas in power-law potentials. In particular, 
we study the role of the inter-atomic 
interaction and finite-size effects. We derive analytical 
formulas for the total number of particles and the 
critical temperature of BEC phase transition. 
\par 
The starting point of our investigation 
is the semiclassical density profile 
$\rho^{(0)}({\bf r})$ of an ideal Bosons in a trapping external 
potential $U({\bf r})$. In the grand canonical ensemble, 
the density profile is given by 
\beq
\rho^{(0)}({\bf r})= \left({mkT\over 2\pi\hbar^2}\right)^{3\over 2} 
B_{3/2}\left(e^{(\mu -U({\bf r}))/kT}\right) \; ,
\eeq 
where $T$ the absolute temperature with 
$k$ the Boltzmann constant, $\mu$ is the chemical potential 
and 
\beq 
B_{r}(z)= \sum_{i=1}^{\infty}{z^i\over i^r} \; ,
\eeq
is the so-called Bose function. 
This result is the generalization of the formula for 
an ideal Bose gas in a box of volume $V$. 
It shows that, in the semiclassical limit, 
the non-homogeneous formula is obtained 
with the substitution $\mu \to \mu - U({\bf r})$, 
also called local density approximation.$^{7,12}$ 
\par 
In the case of weakly-interacting Bosons, the previous formula 
must be substituted by the following one 
\beq
\rho({\bf r})=\left({mkT\over 2\pi\hbar^2}\right)^{3\over 2} 
B_{3/2}\left(e^{(\mu -U({\bf r})+2g \rho({\bf r}))/kT}\right) \; ,
\eeq 
where $g=4\pi\hbar^2a/m$ is the interaction coupling constant 
fixed by the s-wave scattering length $a$. 
This equation is a mean-field self-consistent equation 
that can be derived, within the semiclassical approximation 
and above the critical temperature $T_c$, 
from the Bogolibov-Popov equations, which describe 
the elementary excitations of the Bose gas.$^{4,7}$ 
The chemical potential $\mu$ is fixed by imposing 
the following normalization condition 
\beq 
N= \int \rho({\bf r})\; d^3{\bf r} \; , 
\eeq 
where $N$ is the total number of Bosons. Note that 
below the critical temperature $T_c$, the density 
of the left-hand side in Eq. (3) describes only the 
non-condensed thermal cloud, while the density 
in the right-hand side is the total density, namely 
condensed and non-condensed density. 
Nevertheless, to calculate the BEC transition temperature $T_c$ 
it is sufficient the study the non-condensed thermal cloud. 
In fact, by imposing for the density profile (3) 
the normalization condition (4) one finds the 
total number of particles $N$ as a function of $T$ and $\mu$. 
At the critical temperature $T_c$, the chemical 
potential $\mu$ is minimum ($\mu=0$ in the large $N$ limit)  
and thus one gets the critical temperature $T_c$ 
as an implicit function of $N$.$^{4-7,12}$ 
\par 
Let us consider a generic isotropic 
power-law potential given by  
\beq 
U({\bf r})= A \; r^n = {1\over 2} \left(
{\hbar \omega_0\over r_0^n}\right) \; r^n \; , 
\eeq
where $n$ is the power-law exponent and $A$ is the trap constant. 
Here we have introduced an energy parameter $\hbar \omega_0$  
and a length parameter $r_0$, which can be chosen 
as $r_0=(\hbar/(m\omega_0))^{1/2}$. 
By Taylor expanding the Eq. (3) to the first order in the 
coupling constant $g$ one finds 
\beq 
\rho({\bf r}) = \rho^{(0)}({\bf r}) - 
{2 g\over kT} \left({mkT\over 2\pi\hbar^2}\right)^{3\over 2} 
B_{1/2}\left(e^{(\mu -U({\bf r}))/kT}\right) \rho^{(0)}({\bf r}) \; ,  
\eeq 
where $\rho^{(0)}({\bf r})$ is the non-interacting density 
given by Eq. (1). Then, by imposing the normalization 
condition one gets 
\beq
N = \left({kT\over \hbar \omega_0}\right)^{3(n+2)\over 2n} 
{2^{3\over n} \Gamma({3\over n}+1) \over 2^{3\over 2} 
\Gamma({3\over 2}+1)} 
B_{3(n+2)\over 2n}(e^{\mu/kT})  
\left[ 1 - 2 g \left({m\over 2\pi\hbar^2}\right)^{3\over 2} 
\left( kT \right)^{1\over 2} 
{S_{3\over n}(e^{\mu/kT}) \over B_{3(n+2)\over 2n}
(e^{\mu/kT})}\right] \; ,  
\eeq 
where $\Gamma(r)$ is the Euler factorial function and 
\beq 
S_{r}(z)= \sum_{ij=1}^{\infty} 
{z^{i+j}\over i^{3\over 2} j^{1\over 2} (i+j)^r} \; . 
\eeq 
In Eq. (7) the total number $N$ 
of Bosons is a function of temperature $T$, 
chemical potential $\mu$ and interaction constant $g$,  
namely $N=N(T,\mu,g)$. 
Setting $g=0$ we have the number of particles for an 
ideal Bose gas. In this case, to find the critical temperature 
$T_c^{(0)}$ we put $\mu =0$ and obtain 
\beq 
k T_c^{(0)} = c(n) \; 
\hbar \omega_0 \; N^{2n\over 3(n+2)} \; , 
\eeq
where $c(n)$ is a numerical coefficient,  
\beq
c(n) = 
\left[ {2^{3\over 2}\Gamma({3\over 2}+1) \over 2^{3\over n}
\Gamma({3\over n}+1) 
\zeta({3\over n} + {3\over 2}) } 
\right]^{2 n\over 3(n+2)} \; ,  
\eeq
and $\zeta(r)=B_r(1)$ is the Riemann $\zeta$-function. 
As previously stated, 
this formula was obtained for the first time by 
Bagnato, Pritchard and Kleppner.$^{5}$ 
\par 
To determine the effect of the inter-atomic interaction  
on the critical temperature, we expand the Eq. (7)  
around $\mu=0$, $T=T_c^{(0)}$ and $g=0$: 
$$ 
N(T,\mu,g)=N(T_c^{(0)},0,0)+
{\partial N\over \partial T}(T_c^{(0)},0,0) \; (T-T_c^{(0)}) +
$$
\beq 
+ {\partial N\over \partial \mu}(T_c^{(0)},0,0) \; \mu + 
{\partial N\over \partial g}(T_c^{(0)},0,0) \; g  \; .  
\eeq 
At the critical temperature ($T=T_c$) of the interacting system, 
for large $N$, the chemical potential can be written 
$\mu = 2 g \rho^{(0)}({\bf 0})$ as suggested by Eq. (3).$^{4,7}$  
In this way, from Eq. (1), Eq. (9) and Eq. (11) 
we get the main result of the paper 
\beq 
{ T_c - T_c^{(0)} \over T_c^{(0)} } = - d(n) 
\left( {a\over r_0} \right) N^{n\over 3(n+2)} \; , 
\eeq 
where 
\beq	
d(n) ={ 2^{3\over 2} \left( \zeta({3\over n}+
{1\over 2}) \zeta({3\over 2}) -S({3\over n}) \right) 
\over \pi^{1\over 2} \zeta ({3\over n} +{3\over 2})
({3\over n}+{3\over 2})} 
\left[{ 2^{3\over 2} \Gamma({3\over 2}+1) 
\over 2^{3\over n} \Gamma({3\over n}+1) 
\zeta({3\over n}+{3\over 2}) } \right]^{n\over 3(n+2)} \; ,  
\eeq
and $S(r)=S_r(1)$. 
An inspection of Eq. (12) shows that the shift due to the interaction 
can be either negative or positive, 
depending on the sign of the scattering length $a$ 
and of the coefficient $d(n)$. 
As shown in Tab. 1, $d(n)$ diverges for $n=6$ 
and goes to zero as $n\to \infty$. 
\par
The temperature shift (12) scales as $N^{n\over 3(n+2)}$, where 
$n$ is the exponent of the power-law potential. 
It means that increasing $n$ one can amplify 
the shift of the critical temperature 
due to the interaction. This effect is shown in Fig. 1, 
where we plot the critical temperature $T_c$ 
as a function of the number $N$ of bosons for 
$n=2$ and $n=4$. It is important to observe that in our approach 
the shift is always associated with a change 
in the central density produced by inter-atomic forces.
The relationship between the critical temperature 
and the corresponding value of the density in the 
center of the trap is unaffected by the interaction. 
Moreover, the shift vanishes for a homogeneous 
system where the density is uniform. 
Note that with $n=2$ one 
recovers the shift-formula obtained by Giorgini, Pitaevskii 
and Stringari for a dilute Bose 
gas in harmonic potential.$^{7}$ 
\par 
Another interesting aspect of BEC in trapping potentials 
are finite-size effects. In an ideal gas, 
BEC starts at the temperature 
for which the chemical potential $\mu$ reaches the 
energy of the lowest solution of the Schr\"odinger 
equation. In the case of a power-law potential 
this energy is $\alpha_n \hbar \omega_0$, where 
$\alpha_n$ is a coefficient depending on the 
power-law exponent $n$. For example, in the 
case of the harmonic potential we have 
$\alpha_2=3/2$. In general, the coefficient $\alpha_n$ 
must be numerically evaluated and only 
in the large $N$ limit one can set $\mu=0$ 
at the transition temperature. 
We can estimate finite-size effects on the basis of Eq. (11) 
by setting $g=0$, $\mu = \alpha_n \hbar \omega_0$ and 
$T=T_c^{(FS)}$. In this way we obtain 
\beq 
{T_c^{(FS)} - T_c^{(0)} \over T_c^{(0)} } = 
- e(n) \alpha_n N^{-{2n\over 3(n+2)}} 
\eeq 
where 
\beq
e(n) = { \zeta({3\over n}+{1\over 2}) 
\over \zeta ({3\over n} +{3\over 2})({3\over n}+{3\over 2})} 
\left[{ 2^{3\over 2} \Gamma({3\over 2}+1) \over 2^{3\over n} 
\Gamma({3\over n}+1) 
\zeta({3\over n}+{3\over 2}) } \right]^{-{2n\over 3(n+2)}} \; . 
\eeq 
Numerical values of the coefficient $e(n)$ are reported in Tab. 1. 
The temperature shift (14) originating from finite size effects 
changes sign as at $n=6$, where the coefficient $e(n)$ diverges, 
and, as expected, the shift vanishes in the large $N$ limit. 
In particular, with a fixed number $N$ of Bosons, 
one can strongly reduce this finite-size shift 
by increasing the power-law exponent $n$ as shown in Fig. 1, 
where we have set $\alpha_4=\alpha_2=3/2$. 
The finite-size shift predicted by Grossmann and Holthaus$^{6}$ 
for a Bose gas in harmonic potential is obtained 
from Eq. (14) with $n=2$. 
\par 
In conclusion, we have derived two remarkable formulas 
for the critical temperature of a weakly-interacting Bose gas 
in a generic isotropic power-law potential. 
One formula gives, to the first order in the scattering length, 
the shift of the critical temperature due to 
the inter-atomic interaction. 
The temperature shift is strongly enhanced by a large 
the power-law exponent but the sign of the shift depends 
in a non trivial way on the power-law exponent. 
The other formula shows that, 
as expected, the finite-size effects, namely setting the chemical 
potential equal to zero at the Bose-Einstein 
transition, are negligible in the large $N$ limit. 
Moreover, with a fixed number $N$ of bosons, the finite-size 
effects are reduced by a large power-law exponent. 

\newpage

\section*{References}

\begin{description}

\item{\ 1.} M.H. Anderson, J.R. Ensher, M.R. Matthews, C.E. Wieman, 
and E.A. Cornell, Science 269 (1995) 189. 

\item{\ 2.} K.B. Davis, M.O. Mewes, M.R. Andrews, N.J. van Druten, 
D.S. Drufee, D.M. Kurn, and W. Ketterle, Phys. Rev. Lett. 75 
(1995) 3969. 

\item{\ 3.} C.C. Bradley, C.A. Sackett, J.J. Tollet, and R.G. Hulet, 
Phys. Rev. Lett. 75 (1995) 1687. 

\item{\ 4.} F. Dalfovo, S. Giorgini, L.P. Pitaevskii, and 
S. Stringari, Rev. Mod. Phys. 71 (1999) 463. 

\item{\ 5.} V. Bagnato,  D.E. Pritchard and D. Kleppner, 
Phys. Rev. A 35 (1987) 4354. 

\item{\ 6.} S. Grossmann and M. Holthaus, Phys. Lett. A 
208 (1995) 108. 

\item{\ 7.} S. Giorgini, L. Pitaevskii and S. Stringari, 
Phys. Rev. A 54 (1996) 4633. 

\item{\ 8.} B. Pozzi, L. Salasnich, A. Parola and 
L. Reatto, Eur. Phys. Jour. D 11 (2000) 367. 

\item{\ 9.} W. Ketterle and D.E. Pritchard, 
Phys. Rev. A 46(1992) 4051. 

\item{\ 10.} P.W.H. Pinkse, A. Mosk, M. Weidenmuller, 
M.W. Reynolds, T.W. Hijmans, and J.T.M. Walraven, 
Phys. Rev. Lett. 78 (1997) 990. 

\item{\ 11.} D.M. Stamper-Kurn, H.J. Miesner, A.P. 
Chikkatur, S. Inouye, J. Stenger, and W. Ketterle, 
Phys. Rev. Lett. 81 (1998) 2194. 

\item{\ 12.} L. Salasnich, J. Math. Phys. 41 (2000) 8016. 

\end{description}

\newpage

\begin{center}
\begin{tabular}{|c|cccccccc|} \hline 
$$ & $1$ & $2$ & $3$ & $4$ & $5$ & $6$ & $7$ & $8$  \\ 
\hline 
$c(n)$ & $0.56$ & $0.94$ & $1.14$ & $1.25$ & $1.31$ & $1.35$ & $1.04$ 
& $1.00$ \\
$d(n)$ & $0.64$ & $1.33$ & $-0.23$ & $2.41$ & $7.24$ & $\infty$ 
& $-41.57$ & $-42.96$ \\ 
$e(n)$ & $0.40$ & $0.50$ & $0.29$ & $1.11$ & $2.45$ & $\infty$ 
& $-2.96$ & $-1.61$ \\
${2n\over 3(n+2)}$ & ${2/9}$ & ${1/3}$ & ${2/5}$ 
& ${4/9}$ & ${10/21}$ & ${1/2}$ & ${14/27}$ & ${8/15}$ \\
${n\over 3(n+2)}$ & ${1/9}$ & ${1/6}$ & ${1/5}$ & 
${2/9}$ & ${5/21}$ & ${1/4}$ & ${7/27}$ & ${4/15}$ \\
\hline 
\end{tabular} 
\end{center} 
\par
{TAB. 1}. Numerical coefficients as a function 
of the power-law exponent $n$. 
In the limit $n\to \infty$ one gets: 
$c(n)\to 1.27$, $d(n)\to 0$, $e(n)\to -0.29$, 
${2n/(3(n+2))} \to 2/3$, and ${n/(3(n+2))}\to 1/3$. 

\newpage 

\begin{figure}
\centerline{\psfig{file=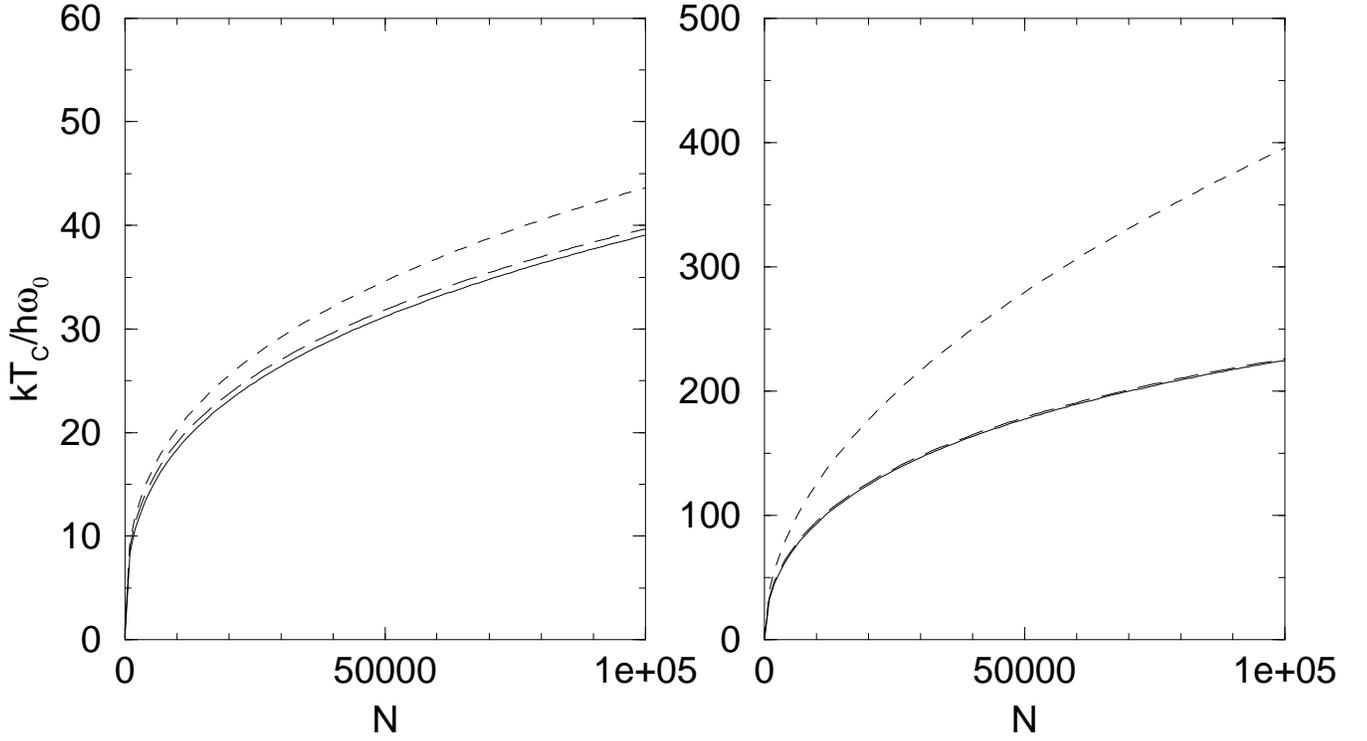,height=4.in}}
\caption{Critical temperature $T_c$ as a function 
of the number $N$ of Bosons. Power-law exponent $n=2$ 
on left and $n=4$ on the right. Ideal Bose gas (dashed line), 
interacting Bose gas (long dashed line), interacting 
Bose gas with finite-size corrections (full line). 
Interaction strength $a/r_0=10^{-2}$.} 
\end{figure}

\end{document}